\documentclass[amsfonts,showpacs,tightenlines,aps,12pt,floatfix]{revtex4}
\usepackage{bm}
\usepackage{epsfig}
\begin{document}


\title{$V_{us}$ From Hadronic $\tau$ Decays}
\author{Kim Maltman}
\affiliation{Department of Mathematics and Statistics, York University, 
4700 Keele
St., Toronto, ON CANADA M3J 1P3}
\altaffiliation{CSSM, Univ. of Adelaide, Adelaide, SA 5005 AUSTRALIA}

\author{Carl E. Wolfe}
\affiliation{Department of Physics and Astronomy, York University, 
4700 Keele
St., Toronto, ON CANADA M3J 1P3}
\date{\today}

\begin{abstract}
We study the reliability of extractions of $\vert V_{us}\vert$ based 
on flavor-breaking hadronic $\tau$ decay sum rules. The 
``$(0,0)$ spectral weight'', proposed previously
as a favorable candidate for this extraction,
is shown to produce results having poor stability
with respect to $s_0$, the upper limit on the relevant spectral integral,
suggesting theoretical errors much larger than 
previously anticipated. We argue that this instability
is due to the poor convergence of the integrated $D=2$ OPE 
series. Alternate weight choices designed
to bring this convergence under better control are shown to produce 
significantly improved stability, and determinations of $\vert V_{us}\vert$ 
which are both mutually compatible, and consistent, within errors,
with values obtained by other methods.
\end{abstract}


\maketitle

\section{\label{intro}Background}
Three-family unitarity of the Cabibbo-Kobayashi-Maskawa (CKM) matrix implies
\begin{equation}
\vert V_{ud}\vert^2 + \vert V_{us}\vert^2 + \vert V_{ub}\vert^2 = 1\ ,
\label{constraint}\end{equation}
with the $V_{ub}$ contribution playing a numerically negligible 
role~\cite{pdg04}. Analyses of $K_{\ell e3}$ incorporating
recent updates to the $K_L$ lifetime~\cite{kloekllifetime},
the $K^+$~\cite{e86503}, $K_L$~\cite{klbrs} and
$K_s$~\cite{kloeks06} branching fractions, and the $K_{\ell 3}$ form 
factor slope parameters~\cite{ke3slopes},
together with strong isospin-breaking and long distance electromagnetic 
corrections computed in the framework of ChPT~\cite{longdistemetc}, 
lead to~\cite{ckmwg1}
\begin{equation}
f_+(0)\vert V_{us}\vert \, =\, 0.2173\pm 0.0008\ ,
\label{ckmwg1result}\end{equation}
which, with the Leutwyler-Roos estimate, 
$f_+(0)=0.961\pm 0.008$~\cite{lr84} (compatible within errors with 
recent quenched and unquenched lattice results~\cite{latticefzero}),
yields~\cite{ckmwg1}
\begin{equation}
\vert V_{us}\vert \, =\, 0.2261\pm 0.0021\ .
\label{vuske3}\end{equation}
This result is in good agreement with expectations based on
unitarity and the most recent update
of the average of superallowed $0^+\rightarrow 0^+$ nuclear $\beta$ 
decay~\cite{hardy04} and neutron decay~\cite{abele02} results,
$\vert V_{ud}\vert =0.9738\pm 0.0003$~\cite{ckmwg1}. The $\sim 2\sigma$ 
discrepancy observed when earlier $K$ decay results were employed
thus appears finally to have been resolved. One should, however, bear in 
mind two recent developments relevant to $\vert V_{ud}\vert$:
(i) a new measurement of the neutron lifetime, in strong 
disagreement with the previous world average~\cite{serebrov04},
and (ii) a Penning trap measurement of the $Q$ value of
the superallowed ${}^{46}V$ decay~\cite{savard05} in significant 
disagreement with the average used as input in Ref.~\cite{hardy04}, 
and with the potential to raise doubts about 
current evaluations of structure-dependent isospin-breaking 
corrections~\cite{hardy05}. The potentially unsettled $\vert V_{ud}\vert$
situation makes alternate (non-$K_{\ell 3}$) determinations of 
$\vert V_{us}\vert$ of interest, both as a means of testing
the Standard Model (SM) scenario for strangeness-changing interactions,
and for reducing errors through averaging. 
Two such alternate methods have been proposed recently.

In the first, $\vert V_{us}/V_{ud}\vert$ is extracted using lattice results 
for $f_K/f_\pi$ in combination with experimental results for 
$\Gamma [K_{\mu 2}] / \Gamma [\pi_{\mu 2}]$~\cite{marciano04}.
With the recently updated MILC $n_f=3$ unquenched lattice result, 
$f_K/f_\pi = 1.198^{+.016}_{-.006}$~\cite{milc05}, the first method yields 
\begin{equation}
\vert V_{us}\vert \, =\, 0.2245^{+0.0011}_{-0.0031}\ ,
\label{marcianovus}\end{equation}
compatible within errors with the $K_{\ell 3}$ determination.

The second of these proposals involves the analysis of flavor-breaking sum 
rules employing strange and non-strange hadronic $\tau$ decay 
data~\cite{pichetalvus}, and forms the subject of the rest of 
this paper. Existing results, based on the ``$(0,0)$ spectral weight'' 
version of this analysis, will be discussed as part of the development below.
The discussion to follow represents an update and extension
of the preliminary results presented in Ref.~\cite{krmtau04}.

\section{\label{section2}$V_{us}$ From Hadronic $\tau$ Decay Data}
With $\Pi^{(J)}_{V/A;ij}$ the spin $J$ parts of the flavor $ij=ud,us$ 
vector/axial vector correlators, $\rho^{(J)}_{V/A;ij}$ the
corresponding spectral functions, and 
$R_{V/A;ij}\, \equiv\, \Gamma [\tau^- \rightarrow \nu_\tau
\, {\rm hadrons}_{V/A;ij}\, (\gamma )]/
\Gamma [\tau^- \rightarrow \nu_\tau e^- {\bar \nu}_e (\gamma)]$,
the kinematics of hadronic $\tau$ decay imply~\cite{bnpetc}
\begin{equation}
R_{V/A;ij}\, =\, 12\pi^2\vert V_{ij}\vert^2 S_{EW}\,
\int^{m_\tau^2}_{th}\, {\frac{ds}{m_\tau^2}} \,
\left( 1-y_\tau\right)^2
\, \left[ \left( 1 + 2y_\tau\right)
\rho_{V/A;ij}^{(0+1)}(s) - 2y_\tau \rho_{V/A;ij}^{(0)}(s) \right]
\label{basictaudecay}\end{equation}
where $y_\tau =s/m_\tau^2$, $V_{ij}$ is the
flavor $ij$ CKM matrix element, $S_{EW}=1.0201\pm 0.0003$~\cite{erler02} 
is a short-distance electroweak correction, and the superscript
$(0+1)$ denotes the sum of $J=0$ and $J=1$ contributions.
Eq.~(\ref{basictaudecay}) is written in such a way
that both terms on the RHS can be rewritten
using the general finite energy sum rule (FESR) relation,
\begin{equation}
\int_{th}^{s_0}ds\, w(s) \rho (s)\, =\, {\frac{-1}{2\pi i}}\,
\oint_{\vert s\vert =s_0}ds\, w(s) \Pi (s)\ ,
\label{fesrbasic}\end{equation}
valid for any analytic weight $w(s)$ and any correlator $\Pi$ without
kinematic singularities. 
Quantities $R_{V/A;ij}^{(k,m)}$, analogous to $R_{V/A;ij}$,
are obtained by rescaling the experimental decay distribution
with the factor $(1-y_\tau )^ky_\tau^m$ before integrating. The corresponding
FESR's are referred to as the ``$(k,m)$ spectral weight sum rules''. 
Similar FESR's can be written down for general weights $w(s)$,
for $s_0<m_\tau^2$, and for the
separate correlator combinations $\Pi^{(0+1)}_{V/A;ij}(s)$ and
$s\Pi^{(0)}_{V/A;ij}(s)$. The corresponding
spectral integrals, $\int_{th}^{s_0}ds\, w(s) \rho^{(J)}_{V/A;ij} (s)$, 
will be denoted $R^w_{ij}(s_0)$ in what follows. 
In FESR's involving both the $J=0+1$ and $J=0$ combinations, 
the purely $J=0$ contribution will be referred to as ``longitudinal''.

With this background, the $\tau$-based extraction of $V_{us}$ works 
schematically as follows~\cite{pichetalvus}. Given 
experimental values for the spectral integrals $R^w_{ij}(s_0)$,
$ij=ud,us$, corresponding to the same $w(s)$ and same
$s_0$, the combination 
\begin{equation}
\delta R^w(s_0)\, =\, {\frac{R^w_{ud}(s_0)}{\vert V_{ud}\vert^2}}
\, -\, {\frac{R^w_{us}(s_0)}{\vert V_{us}\vert^2}}
\label{tauvusbasicidea}\end{equation}
vanishes in the SU(3) flavor limit and hence has an OPE representation,
$\delta R^w_{OPE}(s_0)$, which begins at dimension $D=2$. Solving for
$\vert V_{us}\vert$, one has
\begin{equation}
\vert V_{us}\vert \, =\, \sqrt{ {\frac{R^w_{us}(s_0)}
{\left[ R^w_{ud}(s_0)/\vert V_{ud}\vert^2\right]\, -\, 
\delta R^w_{OPE}(s_0)}}}\ .
\label{tauvussolution}\end{equation}
At scales $\sim 2-3\ {\rm GeV}^2$, and for weights used in the 
literature, the dominant $D=2$ term in $\delta R^w_{OPE}(s_0)$  
is much smaller than the leading $D=0$ contribution and, as a consequence, 
similarly smaller than the separate $ud$, $us$ spectral integrals
(for physical $m_s$, typically at the few to several percent level). 
The OPE uncertainty, $\Delta \left(\delta R^w_{OPE}(s_0)\right)$, 
thus produces a fractional $\vert V_{us}\vert$ error 
$\simeq \Delta \left(\delta R^w_{OPE}(s_0)\right)/2\, R^w_{ud}(s_0)$, 
{\it much} smaller than the fractional uncertainty on 
$\delta R^w_{OPE}(s_0)$ itself. High accuracy for $\vert V_{us}\vert$
is thus obtainable with only modest accuracy for $\delta R^w_{OPE}(s_0)$
provided experimental spectral integral errors can be kept under control.

At present, the absence of a V/A separation of the $us$ spectral data
means one must work with sum rules based on the observed V+A combination.
This combination also reduces the fractional $ud$ spectral integral
errors. With present $ud$ spectral data~\cite{cleoud95,alephud98,opalud99}, 
these errors are at the $\sim 0.5\%$ level for weights used 
previously in the literature. The much smaller strange branching fraction
leads to limited statistics and coarser binning for the $us$ spectral
distribution~\cite{alephus99,cleous0305,opalus04}. 
The $K$ pole term is very accurately known, but errors are $\sim 6-8\%$
in the $K^*$ region and $>20-30\%$ above $1\ {\rm GeV^2}$. For weights used 
in the literature, the result is $us$ spectral integrals
with $\sim 3-4\%$ uncertainties~\cite{alephus99,chen01,opalus04}.
Experimental errors on $\vert V_{us}\vert$ are thus at the
$\sim 1.5-2\%$ level, and dominated by uncertainties in the $us$ sector.
The situation should improve dramatically with the increase in statistics 
and improved $K$ identification available from the B factory experiments.

A number of points relevant to reducing OPE errors are outlined below.
Note that use of the V+A sum rules 
has the added advantage of strongly suppressing duality 
violation at the scales considered~\cite{krmtau02}. Working
with weights satisfying $w(s=s_0)=0$ further suppresses
such contributions~\cite{krmtau02,krmpfesr}, as does working
at scales $s_0>2\ {\rm GeV}^2$~\cite{cdgmudvma}.

A major, and irreducible, source of OPE uncertainty for
``inclusive'' sum rules (those involving both $J=0+1$ and $J=0$ contributions)
is that produced by the bad behavior of the integrated longitudinal 
$D=2$ OPE series. This representation displays badly non-convergent behavior, 
order by order in $\alpha_s$, even at the maximum scale, $s_0=m_\tau^2$, 
allowed by kinematics~\cite{longitudinalconvergence}. Moreover, 
for the $(k,0)$ spectral weights, those truncations of this series employed 
in the literature can be shown to strongly 
violate constraints associated with the positivity of the continuum 
(non-$K$-pole) part of $\rho_{V+A;us}^{(0)}(s)$~\cite{positivityviolation}. 

The impossibility of making sensible use of the longitudinal OPE 
representation necessitates 
working with sum rules based on the $J=0+1$ combination. 
Since no complete $J=0/1$ spin separation of the spectral 
data exists, a phenomenological subtraction of the longitudinal parts 
of the experimental decay distribution is necessary. 
This can be done with good accuracy because the (very accurately known)
$\pi$ and $K$ pole terms dominate the subtraction, for a combination
of chiral and kinematic reasons~\cite{krmtau02,kmms00}.
Small continuum $us$ longitudinal corrections have 
been constrained phenomenologically,
by a sum rule analysis of the flavor $us$ pseudoscalar channel~\cite{mkps01}
and a coupled-channel dispersive analysis of the scalar channel
which employs experimental $K\pi$ phases, ChPT, and short-distance 
QCD constraints as input~\cite{jopss}. 
The contribution of the 
resulting phenomenological longitudinal continuum $us$ model 
to the $\tau$ strangeness branching fraction (corresponding to the 
$(0,0)$ spectral weight, and $s_0=m_\tau^2$) is $<1\%$ of the total. 
The uncertainty in the bin-by-bin continuum longitudinal subtraction 
thus plays a minor role, even for the $(0,0)$ spectral 
weight. Dominance of the longitudinal $us$ continuum by
the $K_0^*(1430)$ and $K(1460)$ resonance contributions also means 
the impact of longitudinal subtraction uncertainties on 
the resulting $J=0+1$ spectral integrals decreases rapidly with 
decreasing $s_0$ and is much reduced for weights (like
the $(k>0,0)$ spectral weights) which fall off to zero faster at $s=s_0$ 
than does the $(0,0)$ spectral weight.

For the remainder of this paper, we focus
on the flavor-breaking combination
\begin{equation}
\Delta\Pi (s)\, \equiv\, \Pi_{V+A;ud}^{(0+1)}(s)\, -\, 
\Pi_{V+A;us}^{(0+1)}(s)\ .
\label{correlatorchoice}\end{equation}

\subsection{OPE Input}
The OPE representation of $\Delta\Pi$ is known up to dimension $D=6$.

The leading, $D=2$, term is given by~\cite{bck05}
\begin{equation}
\left[\Delta\Pi (Q^2)\right]_{D=2}\, =\, {\frac{3}{2\pi^2}}\,
{\frac{\bar{m}_s}{Q^2}} \left[ 1+2.333 \bar{a}+19.933 \bar{a}^2
+208.746 \bar{a}^3+(2378\pm 200)\bar{a}^4+\cdots \right]
\label{d2form}\end{equation}
where $\bar{a}=\alpha_s(Q^2)/\pi$ and $\bar{m}_s=m_s(Q^2)$, with
$m_s(Q^2)$ and $\alpha_s(Q^2)$ the running strange quark
mass and coupling in the $\overline{MS}$ scheme.
The $O(\bar{a}^4)$ term~\cite{bck05} was estimated using methods which
provided an extremely accurate prediction for the $O(\bar{a}^3)$ 
coefficient, and similarly reliable predictions of the $n_f$-dependent 
$O(\bar{a}^3m_q^2)$ coefficients of the electromagnetic current 
correlator~\cite{bckzin04}, all in advance of their explicit calculation.
For $\bar{a}$ and $\bar{m}_s$, we employ exact solutions corresponding to
the 4-loop-truncated $\beta$ and $\gamma$ functions~\cite{betagamma4}, 
with initial conditions for $m_s$~\cite{pdg04} and $\alpha_s$~\cite{alephud98}
\begin{equation}
m_s(2\ {\rm GeV})=105\pm 25\ {\rm MeV},\ 
\alpha_s(m_\tau^2)=0.334\pm 0.022 .
\label{4loopics}\end{equation}

The forms of the $D=4$ and $D=6$ contributions are well known,
and may be found in Ref.~\cite{bnpetc}. The dominant 
$D=4$ contribution is that proportional to the RG-invariant
strange quark condensate. This is evaluated
using ChPT quark mass ratios~\cite{leutwylerqm}, GMOR for
the light quark condensate, and the conventional estimate 
$r_c\equiv \langle m_\ell \bar{\ell}\ell\rangle /
\langle m_s\bar{s}s\rangle = 0.8\pm 0.2$ for the ratio of
the two condensates. $D=6$ contributions are evaluated
using the vacuum saturation approximation (VSA) and 
assigned an error of $\pm 500\%$.
$D>6$ terms are assumed negligible. This assumption can be tested for
self-consistency since, for polynomial weights, 
$w(y)=\sum_m c_m y^m$ (with $y=s/s_0$), 
integrated $D=2m+2$ OPE contributions scale as
$1/s_0^m$. Neglected, but non-negligible, $D>6$ contributions will thus show 
up as unphysical instabilities in the output of a given sum rule 
(in this case, $\vert V_{us}\vert$) with respect to $s_0$.
Since, when $c_m\not= 0$, the $D=2m+2$ contribution to $\delta R^w_{OPE}(s_0)$ 
is unsuppressed by any additional factors of $\alpha_s$, an 
$s_0$-stability test is particularly important for weights, $w(y)$,
having $D>6$ OPE contributions potentially enhanced through
large values for one or more of the $c_m$ with $m>2$. 
The $(2,0)$, $(3,0)$ and $(4,0)$ $J=0+1$ spectral weights,
$w^{(2,0)}(y)=1-2y-2y^2+8y^3-7y^4+2y^5$,
$w^{(3,0)}(y)=1-3y+10y^3-15y^4+9y^5-2y^6$,
and $w^{(4,0)}(y)=1-4y+3y^2+10y^3-25y^4+24y^5-11y^6+2y^7$,
are examples of weights having such large higher order coefficients.

The $D=2$ OPE integrals are evaluated using the CIPT 
prescription~\cite{ciptbasic}, in which the RG-improved expression
for $\Delta\Pi $, or its Adler function 
$\Delta D(Q^2) \, =\, -Q^2 d\Pi (Q^2)/dQ^2$, 
is used point-by-point along the integration contour. To all orders, 
the two versions of the $D=2$ integral are necessarily equal, being 
related by a partial integration. With both $\Delta\Pi$ and
$\Delta D$ truncated at the same order, however, they differ
by terms of higher order in $\bar{a}$. Our central values 
employ the $O(\bar{a}^4)$-truncated RG-improved correlator version 
(for arguments in favor of this choice, see Ref.~\cite{bck05}). 

For the scales employed in this study,
both the magnitude and error of $\delta R^w_{OPE}(s_0)$ are dominated
by the $D=2$ contribution. The $D=2$ error has two important sources. 
The first is an overall scale uncertainty, associated with the error 
on the input strange quark mass, $\bar{m}_s(2\ {\rm GeV})$.
This uncertainty is $\sim 50\%$ for the PDG04 input, Eq.~(\ref{4loopics}), 
but should be reduced considerably by ongoing progress in unquenched 
lattice simulations. The second source of error is the
truncation of the $D=2$ series.

The $D=2$ truncation error is potentially significant because 
the series in Eq.~(\ref{d2form}) is slowly converging near
the spacelike point on the FESR contour. In fact,
with high-scale determinations of $\alpha_s(M_Z)$~\cite{pdg04} 
corresponding to an $n_f=3$ coupling $\bar{a}(Q^2=m_\tau^2)\simeq 0.10-0.11$, 
the last three terms in Eq.~(\ref{d2form}) are actually slowly increasing 
with order at the spacelike point throughout the whole of the kinematically
allowed region. Convergence of the integrated series will thus typically 
be slow for weights which emphasize this part of the contour.
The $(k,0)$ spectral weights, which involve $2+k$ powers of $1-y$,
fall more and more into this category as $k$ is increased,
since, on the contour ($y=e^{i\theta}$), 
$\vert 1-y\vert^{2+k}\propto sin^{2+k}(\theta /2)$ is 
more and more peaked in the spacelike direction.


We will use two different monitors of the convergence of
the integrated $D=2$ series. The first involves the difference
of the truncated correlator and Adler function versions of the series,
the second the stability with respect to $s_0$ of the sum rule output.

For a series with good convergence, the correlator and Adler function versions 
of the truncated sum should be in good agreement and, moreover, show 
improved agreement with increasing truncation order. 
We define $r^w_k(s_0)$ to be the fractional change 
in the relevant integrated order-$k$-truncated $D=2$ sum produced by 
shifting from the correlator to corresponding Adler function version. 
Large values of $\vert r^w_k(s_0)\vert$ and/or an increase of 
$\vert r^w_k(s_0)\vert$ with $k$ then signal slow convergence 
of the $D=2$ series~\cite{warningfootnote}. We will take {\it twice} the sum, 
in quadrature, of the last included term and the difference between the
truncated correlator and Adler function versions of the sum as our
estimate for the $D=2$ truncation error. The resulting estimate
is considerably more conservative than those used previously in the literature.

Regarding $s_0$ stability, a truncated series well-converged at 
$s_0=m_\tau^2$ should remain so for some range of $s_0<m_\tau^2$. 
Since the exact $\delta R^w(s_0)$ would
produce a $\vert V_{us}\vert$ independent of $s_0$, a well-converged 
truncation of the integrated $D=2$ OPE series should produce 
$\vert V_{us}\vert$ values stable over some interval of $s_0$.
If, however, neglected higher order terms are actually important 
at $s_0=m_\tau^2$, they will be even more so at lower scales, 
making the accuracy of the truncated expression even worse at those
scales, and producing an unphysical $s_0$ dependence to 
the extracted $\vert V_{us}\vert$~\cite{convs0stabfootnote}.
The absence of a stability window in $s_0$ thus implies the unreliability of 
the truncated integrated $D=2$ series and/or the importance of neglected,
but non-negligible, higher dimension terms.

\subsection{Data Input}
For the $ud$ data we use the ALEPH spectral distribution and
covariance matrix~\cite{alephud98}, with overall normalization corrected
for the small changes in the $e$, $\mu$, and strangeness branching
fractions subsequent to the original ALEPH publication.

For the $us$ data, we have, unfortunately, been unable to obtain the 
covariance matrix from the OPAL collaboration. The OPAL 
publication~\cite{opalus04} quotes 
correlated $us$ spectral integral errors only for a range 
of the $(k,m)$ spectral weights, and only for $s_0=m_\tau^2$. 
This information is insufficient to allow the errors resulting from other
choices for either $s_0$ or $w(y)$ to be inferred, precluding implementation 
of the crucial $s_0$-stability test, even for the 
$(k,0)$ spectral weights. We have thus chosen to work with the somewhat
older ALEPH data~\cite{alephus99}, whose covariance matrix is
publicly available. The two data sets differ mainly in the values
of a small number of the strange branching fractions, a particularly important
difference being that for $\tau^-\rightarrow K^-\pi^+\pi^-\nu_\tau$.
To take into account the changes in the branching fraction values,
we follow the strategy adopted in Ref.~\cite{alephetal01}. In this approach,
the ALEPH distribution for each mode is rescaled by the ratio of the current 
world average to the ALEPH 1999 branching fraction value. The resulting 
rescaled mode-by-mode distributions are then recombined to form the modified 
total $us$ V+A spectral distribution~\cite{thankstoshaomin}. This scheme 
should be reliable for modes whose rescalings are close to $1$, but is less 
clearly so for those, like $\tau^-\rightarrow K^-\pi^+\pi^-\nu_\tau$,
where this is not the case. 


The $us$ $K$ pole spectral integral contribution is fixed
by the $\Gamma [K_{\mu 2}]$-based SM prediction.
This is done because (i) the SM prediction is compatible with the observed 
$\tau\rightarrow K\nu_\tau$ branching fraction, but 
$\sim 6$ times more precise, and (ii) Eq.~(\ref{tauvussolution}) 
already pre-supposes the validity of the SM mechanism 
for hadronic $\tau$ decays.

There exist three determinations of the branching fraction 
$B\left[ \tau^-\rightarrow K^-\pi^+\pi^-\nu_\tau\right]$, 
the original ALEPH result, $0.214\pm 0.037\pm 0.029\%$, 
and the more recent CLEO and OPAL results,
$0.384\pm 0.014\pm 0.038\%$ and $0.415\pm 0.059\pm 0.031\%$,
respectively. While the CLEO and OPAL results are in good agreement,
the agreement with ALEPH is less compelling. The OPAL $us$ spectral integral 
results employed the three-fold average, $0.330\pm 0.028\%$, in setting 
the overall normalization of the $K^-\pi^+\pi^-$ spectral distribution. 
Results based on this normalization are denoted `ACO' below. 
To stress the sensitivity to apparently minor changes in the
branching fraction values, as well as the importance of improved precision, 
we also present results (denoted `CO' in what follows) corresponding to the 
alternate rescaling, produced by the average, $0.40\%$, of the OPAL and 
CLEO central values. The `ACO'/`CO' branching fraction difference, though 
only $0.07\%$, represents more than $2\%$ of the $\sim 3\%$ 
total strangeness branching fraction and hence has the potential to
shift $\vert V_{us}\vert$ by as much as $\sim 1\%$ ($\sim 0.0020$).

\section{Analysis and Results}
In this section we first discuss the existing $(0,0)$ spectral
weight analysis, then present alternate determinations, 
based on weights with improved $s_0$-stability for $\vert V_{us}\vert$.

\subsection{The $(0,0)$ Spectral Weight Determination of $V_{us}$}
The $(0,0)$ spectral weight has been proposed in the literature
as a particularly favorable case for the $\vert V_{us}\vert$
analysis~\cite{pichetalvus}. 
A key potential advantage is the {\it very} close cancellation 
between the weighted $ud$ and $us$ spectral integrals. This manifests itself
in the OPE representation in suppressed values for the integrated 
$D=2$ OPE series and hence similarly suppressed values for the 
$m_s$-induced $D=2$ scale uncertainty. This scale uncertainty is the dominant 
component of the estimated theoretical error in Ref.~\cite{pichetalvus}, 
being a factor of $\sim 2$ larger than the estimated $D=2$ truncation 
uncertainty, and much larger than any of the other contribution. The 
combined theory error, $\pm 0.0009$~\cite{pichetalvus}, is swamped by the 
current $\pm 0.0033$ experimental error, but, if reliable, would make the 
$(0,0)$ analysis a very favorable one for use in determining 
$\vert V_{us}\vert$ once the much improved $us$ spectral
data from BABAR and BELLE becomes available. Unfortunately,
as we will see below, the theoretical uncertainty on 
$\delta R^{(0,0)}_{OPE}(m_\tau^2)$ almost certainly 
{\it significantly} exceeds the estimate of Refs.~\cite{pichetalvus}.
Indeed, we will argue that the convergence of the integrated
$D=2$ $(0,0)$ spectral weight series is sufficiently bad that the
$(0,0)$ analysis is, in fact, an {\it un}favorable one for the extraction of
$\vert V_{us}\vert$. Fortunately, alternatives exist with significantly 
improved convergence behavior, which allow one to take advantage of
the general approach proposed in Refs.~\cite{pichetalvus}. We
return to these in the next subsection, after elaborating on
the problematic features of the $(0,0)$ analysis.

The $(0,0)$-spectral-weight-based results of Refs.~\cite{pichetalvus}
were obtained using the $O(\bar{a}^3)$-truncated Adler function
version of the integrated $D=2$ OPE series.
The truncation error was estimated
by combining the magnitude of the last ($O(\bar{a}^3)$) included 
term in quadrature with a measure of the residual scale dependence,
the latter obtained by changing the CIPT scale choice $\mu^2=Q^2$ to 
$\mu^2 = \xi Q^2$, with $0.75<\xi <1.5$. The resulting truncation uncertainty 
is $^{+10\%}_{-20\%}$ of the $O(\bar{a}^3)$-truncated Adler 
function sum. The results of Ref.~\cite{bck05}, however, show that the
truncated sum is decreased by $47\%$ if evaluated instead using
the $O(\bar{a}^4)$ integrated correlator form.
This shift is larger by a factor of nearly $2.5$ than the
truncation error estimate of Refs.~\cite{pichetalvus}.

Further evidence for the slow convergence of the integrated
$D=2$ $(0,0)$ series
is provided by the values of $r^{(0,0)}_k(m_\tau^2)$, given in
Table~\ref{table1}. The values are not small in general, 
and grow rapidly with increasing $k$.
This increase raises serious doubts about any truncation error 
estimate based on features of the $O(\bar{a}^4)$-truncated sum,
even one, like ours, which is more numerically conservative than that of
Refs.~\cite{pichetalvus}. 

\begin{table}
\caption{\label{table1}Values of $r^w_k(s_0)$, as defined in the
text, for various weight choices $w(y)$ and $s_0=m_\tau^2$}
\vskip .1in
\begin{tabular}{|lcccc|}
\hline
Weight&\quad $k=1$\quad &\quad $k=2$\quad &\quad $k=3$\quad &\quad $k=4$\quad\\
\hline
$w^{(0,0)}_{J=0+1}$&-0.01&0.06&0.20&0.67\\
\hline
$\hat{w}_{10}$&-0.11&-0.07&-0.05&-0.03\\
$w_{20}$&-0.11&-0.08&-0.05&-0.03\\
$w_{10}$&-0.10&-0.06&-0.03&-0.01\\
\hline
\end{tabular}
\end{table}

A final illustration of the unreliability of the convergence of the
integrated $D=2$, $(0,0)$ spectral weight series is provided by the 
$s_0$-stability results, shown in columns 2 (ACO case) 
and 6 (CO case) of Table~\ref{table2}. No stability window for 
$\vert V_{us}\vert$ is evident in either case, as expected given the 
indications for poor convergence already discussed above.
The level of instability is much larger than the 
total theory error estimated in Refs.~\cite{pichetalvus},
even if we restrict our attention to the upper half of the 
interval displayed in the table.

\begin{table}
\caption{\label{table2}$\vert V_{us}\vert$ as a function
of $s_0$ for various FESR weight choices and the ACO and CO treatments of the 
$us$ data. $s_0$ is given in units of GeV$^2$}
\vskip .1in
\begin{tabular}{|l|c|ccc||c|ccc|}
\hline
$s_0$&$w^{(0,0)}_{ACO}$&$\hat{w}^{ACO}_{10}$&$w^{ACO}_{20}$&$w^{ACO}_{10}$&
$w^{(0,0)}_{CO}$&$\hat{w}^{CO}_{10}$&$w^{CO}_{20}$&$w^{CO}_{10}$\\
\hline
2.35&0.2149&0.2220&0.2243&0.2201&0.2172&0.2236&0.2255&0.2218\\
2.55&0.2167&0.2218&0.2235&0.2203&0.2192&0.2236&0.2250&0.2223\\
2.75&0.2181&0.2218&0.2230&0.2207&0.2207&0.2239&0.2246&0.2229\\
2.95&0.2193&0.2220&0.2227&0.2211&0.2219&0.2243&0.2245&0.2235\\
3.15&0.2202&0.2223&0.2226&0.2216&0.2228&0.2246&0.2246&0.2241\\
\hline
\end{tabular}
\end{table}

\subsection{Alternate Weight Choices}
From Refs.~\cite{bck05,kmms00} it is clear that
the integrated $J=0+1$, $D=2$ OPE series for the $(k,0)$ spectral weights
display rather unfavorable convergence behavior, making such
weights problematic for use in extracting $\vert V_{us}\vert$.
Although $(k,0)$ spectral weights occur frequently in treatments
of hadronic $\tau$ decay data, one should bear in mind that
one of the primary reasons for their introduction, 
namely the possibility of using them in inclusive analyses,
is entirely negated by the necessity of avoiding inclusive analyses,
which follows from the extremely bad behavior of the integrated 
longitudinal OPE representation.

In Ref.~\cite{kmms00}, the possibility of constructing
weights more suitable for use in $J=0+1$ non-inclusive sum rules was
investigated. These weights were chosen to (i) emphasize
contributions from regions of the contour showing improved
convergence for the $D=2$ $\Delta\Pi$ series,
(ii) suppress contributions from the region
of the spectrum above $\sim 1\ {\rm GeV}^2$ where $us$
spectral errors are large, and (iii) control the size of
higher order coefficients which might otherwise enhance
$D>6$ contributions. Three such weights,
$w_{10}$, $\hat{w}_{10}$, and $w_{20}$~\cite{w20footnote},
were constructed, all having profiles on the timelike axis 
intermediate between those of the $J=0+1$, $(0,0)$ and $(2,0)$ spectral weights,
and hence similarly intermediate relative $us$ spectral integral errors.
The much improved $D=2$ convergence, compared to that of the $(k,0)$ spectral 
weights, is shown explicitly in Ref.~\cite{kmms00}. 
Further evidence for this improvement is contained in 
Table~\ref{table1}, which shows good agreement, improving
with increasing truncation order, between the correlator 
and Adler function versions of the truncated $D=2$ sums.
The contrast to the $(0,0)$ spectral weight case is striking.

As $s_0$ is decreased, terms in the integrated $D=2$ series of $O(\bar{a}^k)$, 
with $k>4$, grow in size relative to the leading $0^{th}$ order term. 
This growth is least rapid for $w_{20}$ and 
most rapid for $w_{10}$. Though the coefficients multiplying these terms in 
$\Delta \Pi$ are not known, this nonetheless indicates that stability 
for the improved $D=2$ convergence will be best for $w_{20}$ and
worst for $w_{10}$. The $r^{w_{10}}_k(s_0)$ values
in fact display a cross-over in sign and increase in magnitude
with increasing $k\leq 4$ below $s_0\sim 2.55\ {\rm GeV^2}$, 
signalling probable deteriorating convergence.
We thus base our final results on the highest available
scale, $s_0=m_\tau^2$, and favor the $\hat{w}_{10}$ and $w_{20}$ 
analyses over that based on $w_{10}$.

\subsection{Results}
Much improved $s_0$ stability is observed for the
weights with improved $D=2$ convergence, particularly 
$w_{20}$ and $\hat{w}_{10}$. Contributions to the
errors on $\vert V_{us}\vert$ for the various weights are
given in Table~\ref{table3}~\cite{datafootnote}. Sources contributing 
$< 0.0003$ theoretical uncertainty for all weights
considered are not listed explicitly but are included in
the total theoretical error. Since the results
for our favored weights, $w_{20}$ and $\hat{w}_{10}$, are
in excellent agreement, and the combined errors are minimized
for the latter, we take as our final determination the
$\hat{w}_{10}$ results. Displaying the larger of the asymmetric
theory errors, the ACO (CO) $us$ data treatments yield the following
results, both compatible, within errors, with those of 
Eqs.~(\ref{vuske3}) and ~(\ref{marcianovus}):
\begin{equation}
\vert V_{us}\vert\, =\, 0.2223\, (0.2246)\, \pm 0.0032_{exp}\pm 0.0038_{th}\ .
\label{acocoresults}\end{equation}

\begin{table}
\caption{\label{table3}Contributions to the error
on $\vert V_{us}\vert$, at $s_0=m_\tau^2$, for various FESR weight choices.}
\vskip .1in
\begin{tabular}{|l|c|c|cccc|c|}
\hline
Weight&$us$ data&$ud$ data&$m_s$-scale&$\delta\alpha_s$&$\delta r_c$&
$D=2$ truncation&Theory (total)\\
\hline
$w^{(0,0)}$&$\pm 0.0040$&$\pm 0.0006$
&${}^{+0.0006}_{-0.0004}$&${}^{+0.0006}_{-0.0007}$&
$\pm 0.0000$&$\pm 0.0020$&$\pm 0.0022$\\
$\hat{w}_{10}$&$\pm 0.0031$&$\pm 0.0006$
&${}^{+0.0036}_{-0.0027}$&${}^{+0.0001}_{-0.0002}$&
$\pm 0.0005$&$\pm 0.0009$&${}^{+0.0038}_{-0.0029}$\\
${w}_{20}$&$\pm 0.0028$&$\pm 0.0007$
&${}^{+0.0051}_{-0.0038}$&${}^{+0.0001}_{-0.0003}$&
$\pm 0.0008$&$\pm 0.0014$&${}^{+0.0054}_{-0.0041}$\\
${w}_{10}$&$\pm 0.0033$&$\pm 0.0006$
&${}^{+0.0028}_{-0.0021}$&${}^{-0.0001}_{-0.0001}$&
$\pm 0.0004$&$\pm 0.0004$&${}^{+0.0028}_{-0.0022}$\\
\hline
\end{tabular}
\end{table}

\section{Conclusions}
We have shown that the values of $\vert V_{us}\vert$ extracted using 
the $(0,0)$ spectral weight sum rule display a sizeable instability 
with respect to $s_0$. This instability, combined with the results of 
Ref.~\cite{bck05}, strongly suggests that the true $D=2$ truncation 
uncertainty is much larger than previously estimated. 
Given the level of instability, even our much more conservative 
estimate, shown in row 1 of Table~\ref{table3}, 
seems far from being overly conservative. We see no 
plausible way of obtaining a reliable, but more restrictive, estimate of this 
uncertainty. The $D=2$ truncation error thus
represents a sizeable, and irreducible, limitation on the 
accuracy of the $(0,0)$ spectral weight determination of $\vert V_{us}\vert$.

In contrast, for the weights $\hat{w}_{10}$, $w_{20}$ and $w_{10}$
(whose integrated $D=2$ series, by design, display improved convergence),
good consistency, and much improved $s_0$-stability,
is found. At present the theoretical errors on $\vert V_{us}\vert$ 
for these weights, shown in Table~\ref{table3}, are dominated by the 
$m_s$-scale uncertainty. Near-term improvements in unquenched lattice 
simulations should significantly reduce this error. A determination of 
$m_s(2\ {\rm GeV})$ to $\pm 5\ {\rm MeV}$, for example, would reduce 
the combined $\hat{w}_{10}$ theory error to $\pm 0.0013$, bringing a 
sub-$1\%$ determination of $\vert V_{us}\vert$ easily within reach 
with the improved B factory data. Note that $\hat{w}_{10}$ would be
favored over $w_{20}$ because of its smaller truncation error.

It is also possible to construct weights having, simultaneously, improved
$D=2$ convergence and reduced $m_s$-scale sensitivity. We have
generated a number of such weights, but find they typically weight
the region of the spectrum above $1\ {\rm GeV}^2$ more strongly
than do those weights discussed above. As a result, with current $us$ data,
they produce very large experimental errors, and are not presently
useful. We will report on these weights 
elsewhere, once the improved $us$ B factory data has become available 
and meaningful stability tests can be performed.

\begin{acknowledgments}
The ongoing support of the Natural Sciences and Engineering Council
of Canada is gratefully acknowledged.
\end{acknowledgments}


\begin{thebibliography}{99}
\bibitem{warningfootnote}A slowly converging series may,
despite its slow convergence, display neither of these problems.
The $(4,0)$ spectral weight case provides an example.
The integrated $D=2$ series is clearly slowly converging since all terms of 
order $1$ through $4$, in both the correlator and Adler
function versions, are similar in size~\cite{bck05}. 
The problem is, however, not exposed by the behavior
of the series of $r^{(4,0)}_k(m_\tau^2)$ values
($-0.17$, $-0.12$, $-0.07$ and $-0.06$ for $k=1,\cdots ,4$), 
though it does show up in the poor $s_0$-stability of the extracted 
$\vert V_{us}\vert$ results (see, for example, the results quoted 
in Ref.~\cite{krmtau04}).
\bibitem{convs0stabfootnote}For example, with $n_f=3$, 4-loop running 
implies $a(2\ {\rm GeV}^2)/a(m_\tau^2)\simeq 1.16$. The 
role of $O(\bar{a}^5)$ and higher terms, relative to that of the
leading $0^{th}$ order term, can thus easily be enhanced by factors of $2$
or more at such lower scales.
\bibitem{w20footnote}$w_{20}(y)$ is given in somewhat
schematic form in Ref.~\cite{kmms00}. Higher precision versions of
the polynomial coefficients are available from the current authors, on request.
\bibitem{datafootnote}Our error for the $(0,0)$ spectral weight is
based on the ALEPH data and hence somewhat larger than that quoted in the 
second of Refs.~\cite{pichetalvus}, which was based on the OPAL data.
\end{thebibliography}
\end{document}